\documentclass[a4paper,12pt]{article}
\usepackage{epsfig,amssymb,amsmath}
\usepackage{graphicx}
\usepackage{amsfonts}

\begin{document}

\newcommand{\be}{\begin{equation}}
\newcommand{\ee}{\end{equation}}
\newcommand{\bea}{\begin{eqnarray}}
\newcommand{\eea}{\end{eqnarray}}
\newcommand{\bean}{\begin{eqnarray*}}
\newcommand{\eean}{\end{eqnarray*}}
\font\upright=cmu10 scaled\magstep1
\font\sans=cmss12
\newcommand{\ssf}{\sans}
\newcommand{\stroke}{\vrule height8pt width0.4pt depth-0.1pt}
\newcommand{\Z}{\hbox{\upright\rlap{\ssf Z}\kern 2.7pt {\ssf Z}}}
\newcommand{\ZZ}{\Z\hskip -10pt \Z_2}
\newcommand{\C}{{\rlap{\upright\rlap{C}\kern 3.8pt\stroke}\phantom{C}}}
\newcommand{\R}{\hbox{\upright\rlap{I}\kern 1.7pt R}}
\newcommand{\HH}{\hbox{\upright\rlap{I}\kern 1.7pt H}}
\newcommand{\CP}{\hbox{\C{\upright\rlap{I}\kern 1.5pt P}}}
\newcommand{\identity}{{\upright\rlap{1}\kern 2.0pt 1}}
\newcommand{\half}{\frac{1}{2}}
\newcommand{\quart}{\frac{1}{4}}
\newcommand{\pr}{\partial}
\newcommand{\bm}{\boldmath}
\newcommand{\I}{{\cal I}} 
\newcommand{\M}{{\cal M}}
\newcommand{\N}{{\cal N}}
\newcommand{\e}{\varepsilon}

\thispagestyle{empty}
\vskip 3em
\begin{center}
{{\bf \LARGE Antikink-Kink Forces Revisited}} 
\\[15mm]

{\bf \large N.~S. Manton\footnote{email: N.S.Manton@damtp.cam.ac.uk}} \\[20pt]

\vskip 1em
{\it 
Department of Applied Mathematics and Theoretical Physics,\\
University of Cambridge, \\
Wilberforce Road, Cambridge CB3 0WA, U.K.}
\vspace{12mm}

\abstract
{We recalculate the force exerted by an antikink on a kink when their
overlapping tail fields are close to either a quadratic or quartic
minimum of the field theory potential. Our uniform method of
calculation exploits the modified Bogomolny equation satisfied by an
accelerating kink. This method has been used before in special
cases, but is shown here to have broad applicability.
}

\end{center}

\vskip 150pt
\leftline{Keywords: Kinks, Solitons, Scalar Field Theory, Forces}
\vskip 1em

\vfill
\newpage
\setcounter{page}{1}
\renewcommand{\thefootnote}{\arabic{footnote}}

\section{Introduction} 

Topological solitons in Lorentz-invariant field theories provide consistent
models for particles \cite{Raj,book,Wei,Shn}. They are typically
smooth and have a finite size and finite energy, so they model
particles as extended objects without the singularities characteristic of
point-like particles acting as sources for fields. Topological
solitons behave in a more satisfactory way than the old, extended
models of electrons and other charged particles, pioneered by
Abraham--Lorentz and Poincar\'e, where the only field is the
electromagnetic field, and the particle is stabilised by some
mechanical stress \cite{Tep, Roh, Bra}. 

When two or more of these solitons are well separated, they
behave as approximately point-like, and one can seek to calculate the force
acting between them. To find this, one supposes that the solitons are
at rest or slowly moving, and treats the force non-relativistically,
as a Newtonian concept. Each soliton is assumed to move
coherently, negligibly affected by radiation fields that
might produce a back-reaction. In more detail, one assumes that
the soliton has some time-dependent centre, together with a
definite shape around the centre that is minimally deformed
by the presence of other solitons. The goal is to find the leading-order
dependence of the force on the inter-soliton separation, for
large separations, as this is insensitive to the
precise definition of where the centres are.

In what follows we shall focus on kink solitons in one spatial
dimension, and calculate the force between an antikink and a kink.
This force is well known for kinks with exponentially-decaying
(short-range) tails, as occur, for example, in the double-well
$\phi^4$-theory \cite{Ma5}. Here, the tails can be linearly
superposed, and the force calculated using the energy-momentum
tensor to find the rate of change of momentum of either the kink or
antikink. The force for kinks with long-range tails, as
occur when one of the minima of the field theory potential $V(\phi)$ is
quartic or higher-order, is less easy to calculate. Various approaches
have been developed \cite{GE,MGGL,NM,Chr,Chr2,dOr,CM}, but
the most reliable appears to depend on making an ansatz for a
rigidly accelerating kink on a half-line, and the corresponding
ansatz for the oppositely accelerating antikink, and requiring
the complete field configuration to have a continuous slope. This ansatz led
to a prediction for the antikink-kink force in a particular scalar
field theory having an octic potential with a single quartic
minimum \cite{NM}. Here, the kink tail falls off inversely with the
distance from the kink centre and the force falls off with the inverse
fourth power of the antikink-kink separation. The coefficient of the inverse
fourth power was obtained by an approximate treatment of the equation for
the accelerating kink, and this was subsequently validated by an
accurate numerical solution \cite{dOr}.
The result was also tested by numerical simulation of the field theory
evolution, and successfully generalised to certain field theories
with 10th- and 12th-order potentials, where the kink tails fall off
even more slowly \cite{Chr2}. For a recent review, see ref.\cite{KS}.

In this note, we revisit this calculation, and find a more universal
version of the inverse-quartic force when the potential
has at least one quartic minimum. This allows us to discuss more
examples than in \cite{NM}. We also show that the same method
reproduces the well-known antikink-kink force when the potential has
a quadratic minimum and the kink tail is short-range.

An accelerating kink acquires a velocity, of course, but the effect of
the velocity is primarily to produce a Lorentz contraction of
the kink. Provided the velocity is small this effect can be neglected
when calculating the acceleration. The force on the kink due to the
antikink is simply the mass of the kink (the mass that the kink would
have in the absence of the antikink) times the kink acceleration. It turns
out that there is a simpler expression for the force than for
the acceleration. 

Most of these calculations rely on some approximations, and the results
cannot be regarded as rigorous. However, in section 6
the equation describing a rigidly accelerating kink, together with
its boundary conditions, is shown to reduce to a mathematically
well-posed, first-order time-independent differential equation that
is a modification of the usual Bogomolny equation describing a kink at
rest. If this equation is solved precisely (by numerically means) then
one could derive an unambiguous relation between the acceleration and the
antikink-kink separation, even at relatively small separations. This
calculation is outlined here but not pursued.

As an antikink and kink approach each other, there is often an 
interesting annihilation process. Annihilation is sometimes quick,
but sometimes follows an extended sequence of kink-antikink
bounces. The annihilation must usually be studied numerically, as it
has chaotic features. Still, it is useful
to have a theoretical understanding of the force and expected
motion of the kink and antikink before annihilation, because when the
separation is large and the force is small, it can be essentially
impossible to determine the force purely numerically. The theoretical
result makes it possible to start numerical simulations at more modest
separations than otherwise. 

\section{Calculating the Antikink-Kink Force}

We consider the standard, classical field theory for a single real scalar
field $\phi(x,t)$ in one spatial dimension, with Lagrangian
\be
L = \int_{-\infty}^{\infty} \left\{
\half \left(\frac{\pr\phi}{\pr t}\right)^2
- \half \left(\frac{\pr\phi}{\pr x}\right)^2
- V(\phi) \right\} dx \,.
\ee
We assume the potential $V$ is smooth, that $V \ge 0$ for all
$\phi$, and that $V$ takes minimal, vacuum values $V=0$ at
$\phi_-$ and $\phi_+$, with $\phi_+ > \phi_-$. Further, we assume that
$V > 0$ between $\phi_-$ and $\phi_+$ and that $V$ has a single
maximum in this range. With these assumptions, there is a static kink
solution interpolating between $\phi_-$ as $x \to -\infty$ and
$\phi_+$ as $x \to \infty$. $V$ may have further vacua outside this
range, but they will not concern us.

The field equation is
\be
\frac{\pr^2\phi}{\pr t^2} - \frac{\pr^2\phi}{\pr x^2}
+ \frac{dV}{d\phi} = 0 \,,
\ee
so the static kink satisfies
\be
\frac{d^2\phi}{dx^2} - \frac{dV}{d\phi} = 0 \,.
\label{staticeq}
\ee
Multiplying by $\frac{d\phi}{dx}$ and integrating, we reduce this
to the first-order equation
\be
\frac{d\phi}{dx} = \sqrt{2V(\phi)} \,,
\label{Bogo0}
\ee
where we choose the non-negative square root, and have imposed the
boundary conditions $\phi(\pm\infty) = \phi_{\pm}$ to fix the
constant of integration. The same equation emerges from the
Bogomolny rearrangement of the energy \cite{Bog}. The formal solution is
\be
\int^{\phi}_{} \frac{d\phi}{\sqrt{2V(\phi)}} = x \,,
\label{formal}
\ee
implicitly defining the kink $\phi(x)$. Varying the lower
limit of integration allows a spatial translation of the kink.  

If we choose the negative square root, the solution is the antikink,
satisfying the boundary conditions $\phi(\pm\infty) = \phi_{\mp}$.
The antikink is related to the kink by a reflection in $x$.

It is useful to have a precise definition for the centre of the
kink, even though the kink is an extended object and generally has no
symmetry under a reflection. We define the centre $X_0$ to be
where $\frac{d\phi}{dx}$ is maximal. Here the kink profile
$\phi(x)$ has its single inflection point, and $V$ takes its maximal
value. We use the notation $\phi_{\rm K}(x)$ to denote the kink
centred at the origin. The kink centred at $X_0$ is $\phi_{\rm K}(x-X_0)$.

As $V \ge 0$, it is convenient to express $V(\phi) = \half U(\phi)^2$,
with $U \ge 0$, and further to introduce $W(\phi)$ such that
$U = \frac{dW}{d\phi}$. Starting from $V$ or $U$, the function
$W$ has an arbitrary additive constant, and we fix
this so that $W(\phi_+) = 0$. $W$ is then strictly negative and
monotonically increasing in the range $\phi_- \le \phi < \phi_+$.
Equation (\ref{Bogo0}) becomes the Bogomolny equation  
\be
\frac{d\phi}{dx} = \sqrt{2V(\phi)} = U(\phi) =\frac{dW}{d\phi} \,.
\label{Bogo}
\ee

The energy of the kink in terms of $W$ is the integral
\be
E = \half \int_{-\infty}^{\infty} \left\{ \left( \frac{d\phi}{dx}
  \right)^2 + \left( \frac{dW}{d\phi} \right)^2 \right\} dx \,.
\ee
After completing the square and using the Bogomolny equation,
this reduces to
\be
E = \int_{-\infty}^{\infty} \frac{d\phi}{dx} \frac{dW}{d\phi} \, dx 
= W(\phi_+) - W_(\phi_-) \,,
\ee
and as the kink is static, the energy $E$ equals the kink mass $M$.
Since we have set $W(\phi_+) = 0$, the kink mass is $M = -W(\phi_-)$, a
positive quantity. The antikink has the same mass.

We now make a further restriction on the potential $V$ and assume
that $\phi_- = 0$. The kink mass is then $M = -W(0)$. By a shift
of $\phi$, if necessary, this restriction can always be achieved, and
although such a shift affects the kink and antikink solutions, it
does not affect the force between them. We also specify more
precisely the nature of the minimum of $V$ at $\phi = 0$, and
assume that it is either quadratic or quartic, so that near
$\phi = 0$,
\be
V(\phi) = \begin{cases} \half\mu^2\phi^2 + o(\phi^2) \qquad
{\rm quadratic \ minimum \,,} \\
\half \mu^2\phi^4 + o(\phi^4) \qquad {\rm quartic \ minimum \,.}
\end{cases}
\ee

Our interest is in a kink on the half-line $x>0$ interacting with an
antikink on the half-line $x<0$. The overlapping kink and antikink
tails have field values close to $\phi = 0$. Experience with
numerical simulations
shows that it is important that $\phi$ is continuous at $x=0$,
and also that the slope $\frac{\pr\phi}{\pr x}$ approaches zero on
either side of $x=0$. Suitable field configurations can be assumed to
be symmetric under spatial reflection in $x$, so these properties hold
for all time. 

The ansatz we use to model a moving, accelerating kink to the
right of $x=0$, is $\phi(x,t) = \chi(x - A(t))$, a slightly modified version
of the static kink, with $A(t) \gg 0$ the moving centre. The velocity $\dot
A$ and the acceleration $\ddot A$ are assumed to be
small. Substituting this ansatz into the field equation, one finds a
term quadratic in $\dot A$ and a term linear in $\ddot A$. We ignore
the first of these, as it just leads to a small Lorentz contraction
of the kink, and denote the instantaneous acceleration of the kink by
$\ddot A = -a$, where $a$ is positive because of the attraction of the
antikink. The result is the accelerating kink equation
\be
\frac{d^2\chi}{dx^2} - a\frac{d\chi}{dx} - \frac{dV(\chi)}{d\chi} = 0 \,,
\label{accelkinkeq}
\ee
which holds for $x > 0$. The boundary conditions are $\frac{d\chi}{dx}
= 0$ at $x=0$, and $\chi \to \phi_+$ as $x \to \infty$. We seek an 
instantaneous solution with $A$ large and $a$ small. (It is
instantaneous because $a$ itself varies as the antikink-kink 
separation varies.)

To proceed, it is sufficient to make some approximations. We first
note that if $a=0$, the equation is that for the static kink,
eq.(\ref{staticeq}), which reduces to the Bogomolny equation (\ref{Bogo}).
In the term with coefficient $a$ in eq.(\ref{accelkinkeq}) we therefore
replace $\frac{d\chi}{dx}$ by $\frac{dW(\chi)}{d\chi}$, with $W$
related to $V$ as before. This seems adequate, as the formal error
is $O(a^2)$, and the derivative of the static kink $\phi_{\rm K}(x-A)$
is close to zero at $x=0$ provided $A \gg 0$. After this
replacement, eq.(\ref{accelkinkeq}) becomes
\be
\frac{d^2\chi}{dx^2} - a\frac{dW(\chi)}{d\chi} -
\frac{dV(\chi)}{d\chi} = 0 \,,
\ee
which can be rewritten (in terms of $\phi$ rather than $\chi$) as
\be
\frac{d^2\phi}{dx^2} - \frac{d{\widetilde V}}{d\phi} = 0 \,,
\label{modifstatic}
\ee
a variant of eq.(\ref{staticeq}), with modified potential
${\widetilde V}(\phi) = V(\phi) + aW(\phi)$. Note that ${\widetilde V}$,
like $V$, is zero at $\phi_+$, and it has a further simple 
zero near the minimum of $V$ at $\phi = 0$, but
shifted to a positive value $\widetilde{\phi}_-$ because $W$ is
negative. The solution of (\ref{modifstatic}) is a modified kink
interpolating between $\widetilde{\phi}_-$ and $\phi_+$ over the
spatial half-line $x \ge 0$, with $\frac{d\phi}{dx} = 0$ at
both endpoints.

Equation (\ref{modifstatic}) reduces to the
modified first-order equation
\be
\frac{d\phi}{dx} = \sqrt{2{\widetilde V}}
\label{modifBogo}
\ee
with formal solution 
\be
\int^{\phi}_{} \frac{d\phi}{\sqrt{2{\widetilde V}(\phi)}} = x \,,
\ee
but this solution is not straightforward to make explicit, as the
square root of ${\widetilde V} = V + aW$ is generally algebraically
complicated. Instead we make a further approximation, and solve this
modified Bogomolny equation only in the neighbourhood of $\phi = 0$,
exploiting the leading monomial form of $V$, and treating $W$ as
having the constant value $W(0) = -M$. This gives the kink a modified
tail to the left of its centre -- a tail of finite length. The remainder
of the kink is approximated by the unmodified static solution.
Ignoring $aW(\phi)$ here seems to just shift the kink centre by
order $a$, which will have only a subleading effect on the relation
between the acceleration and antikink-kink separation.

The calculation is conceptually the same, whether applied to a
potential $V(\phi)$ with a quadratic or quartic minimum at $\phi=0$, but
the details are different, so we discuss these cases separately.

\section{Quadratic Minimum}

In the quadratic case, the modified potential is approximated by
\be
{\widetilde V}(\phi) = \half\mu^2\phi^2 - Ma
\label{Vmodquad}
\ee
in the kink tail region where $\phi \simeq 0$, and the solution of
eq.(\ref{modifBogo}) is 
\be
\int_{\sqrt{\frac{2Ma}{\mu^2}}}^\Phi
\frac{d\phi}{\sqrt{\mu^2\phi^2 - 2Ma}} = X \,.
\ee
The lower limit of integration is where ${\widetilde V}$ is zero,
and the upper limit $\Phi$ is assumed to be of order 1 but small,
so the quadratic approximation to $V$ is still valid. $X$ is the
location to the left of the kink centre where the modified kink
field has value $\Phi$. The choice of $\Phi$ is somewhat arbitrary, so
neither $\Phi$ nor $X$ have much significance.

In the integral we change variables, by setting
$\phi = \sqrt{\frac{2Ma}{\mu^2}} \cosh u$. The lower limit is $u = 0$
and the upper limit is
$\cosh^{-1}\left(\sqrt{\frac{\mu^2}{2Ma}} \, \Phi \right)$,
which is large as $a$ is small. The result is
\be
X = \frac{1}{\mu} \cosh^{-1} \left(\sqrt{\frac{\mu^2}{2Ma}} \, \Phi
\right) \,,
\ee
and approximating $\cosh u$ by $\half e^u$, this simplifies to
\be
X = \frac{1}{\mu} \log \left( \sqrt{\frac{2\mu^2}{Ma}} \, \Phi \right)
\ee
or equivalently
\be
Ma = 2\mu^2\Phi^2 e^{-2\mu X} \,.
\label{forcequad}
\ee

We now match this relation between $\Phi$ and $X$ to the tail solution
for the unmodified kink centred at $X_0$, where $X_0 > X$. This
tail obeys the linearised equation $\frac{d^2\phi}{dx^2} = \mu^2\phi$, so
the solution is of the form $\phi(x) = {\cal A}e^{\mu(x-X_0)}$, and therefore
$\Phi = {\cal A}e^{\mu(X-X_0)}$. Substituting in (\ref{forcequad}), we obtain
\be
Ma = 2\mu^2 {\cal A}^2 e^{-2\mu X_0} \,.
\ee
Here, the amplitude ${\cal A}$ must be determined from the global
kink solution, using the full potential $V(\phi)$.
We see that the force $F=Ma$ that the antikink exerts on the kink has a simpler
expression than the acceleration $a$, since $M = -W(0)$ is
sometimes algebraically complicated or intractible. $2X_0$ can be
identified with the antikink-kink separation $s$, so finally
\be
F = 2\mu^2 {\cal A}^2 e^{-\mu s} \,.
\ee
This is the general form of the antikink-kink force when the potential
minimum at $\phi = 0$ is quadratic and the kink tail is short-range.

In the familiar example of the $\phi^4$ kink, the potential is $V(\phi) =
\half (1 - \phi^2)^2$. Shifting $\phi$ by 1 so that the left-hand
zero of $V$ is at $\phi = 0$ gives $V(\phi) = \half (2-\phi)^2
\phi^2$, so $\mu = 2$. The global kink solution (after shifting
$\phi$ and centering at $X_0$) is
\be
\phi_{\rm K}(x - X_0) = 1 + \tanh(x - X_0) \simeq 2e^{2(x-X_0)} \,,
\ee
where the second expression is the left-hand tail field for $x \ll X_0$.
Therefore ${\cal A} = 2$, and the antikink-kink force becomes
\be
F = 32 e^{-2s} \,,
\ee
a well-known result \cite{Ma5}.

\section{Quartic Minimum}

We now consider the quartic case. The analogue of (\ref{Vmodquad}) is
\be
{\widetilde V}(\phi) = \half\mu^2\phi^4 - Ma
\ee
in the kink tail region. The solution of the first-order
eq.(\ref{modifBogo}) is therefore
\be
\int_{\left( \frac{2Ma}{\mu^2} \right)^{1/4}}^\Phi
\frac{d\phi}{\sqrt{\mu^2\phi^4 - 2Ma}} = X \,.
\label{quartintegral}
\ee
The lower limit of integration is again where ${\widetilde V}$ is zero,
and the somewhat arbitrary upper limit $\Phi$ is of order 1 but small, so
the quartic approximation to $V$ is still valid. $X$ is where the
field value is $\Phi$.

We rescale by setting $\phi = \left( \frac{2Ma}{\mu^2} \right)^{1/4}\psi$.
Then
\be
X = \left( \frac{1}{2Ma\mu^2} \right)^{1/4}
\int_1^{\left( \frac{\mu^2}{2Ma} \right)^{1/4}\Phi}
\frac{d\psi}{\sqrt{\psi^4 - 1}} \,,
\ee
where the upper limit is large, as $a$ is small.
For large $\psi$ we can approximate the integrand by $\frac{1}{\psi^2}$. The
integral is then the complete elliptic integral \cite{GR}
\be
J \equiv K(i) = \int_1^\infty \frac{d\psi}{\sqrt{\psi^4 - 1}}
= \frac{\Gamma\left( \frac{1}{4} \right)^2}
{4\sqrt{2\pi}} \simeq 1.311 \,,
\ee
with a small correction at the upper limit, giving the result
\be
X = \frac{J}{(2Ma\mu^2)^{1/4}} - \frac{1}{\mu\Phi} \,.
\ee
This can be rewritten as
\be
Ma = \frac{1}{2\mu^2} \frac{J^4}{ \left( X + \frac{1}{\mu\Phi}
  \right)^4} \,.
\label{forcequart}
\ee

Next we note that as $V(\phi) \simeq \half \mu^2 \phi^4$ in the tail
region, the first-order equation for the kink tail (in the absence of
the antikink) is
\be
\frac{d\phi}{dx} = \mu\phi^2 \,,
\ee
whose solution is the long-range tail field
\be
\phi(x) = -\frac{1}{\mu(x- \cal{X})} \,,
\label{quartkinktail}
\ee
with $\cal{X}$ a constant determined by the global kink solution.
$\cal{X}$ plays a similar role for a potential with a
quartic minimum as the amplitude $\cal{A}$ does for a potential with
a quadratic minimum. This tail solution implies that
\be
X + \frac{1}{\mu\Phi} = \cal{X} \,.
\ee
Equation (\ref{forcequart}) therefore simplifies to
\be
Ma = \frac{1}{2\mu^2} \frac{J^4}{{\cal{X}}^4} \,.
\label{forcetilde}
\ee
$\cal{X}$ is not the kink centre $X_0$, but typically differs little
from it. If we ignore this difference for large $X_0$,
then we can replace $\cal{X}$ by $X_0$, and in terms of the
separation $s = 2X_0$,
\be
F = Ma = \frac{8}{\mu^2} \frac{J^4}{s^4} \,.
\ee
This is the general, leading form of the antikink-kink force when the
potential minimum at $\phi = 0$ is quartic. Again, it is the force
rather than the acceleration that has the simpler expression.
$\mu$ is easily determined from the detailed form of $V(\phi)$.

\section{Examples with Quartic Minima}

In general, for a potential with a quartic minimum, there is no precisely
defined difference between $X_0$ and $\cal{X}$, because of a logarithmic
correction to the kink tail. However, in special cases, the
logarithmic term is absent and we can be more definite.

Our first example is the symmetric octic polynomial
\be
V(\phi) = \half (1 - \phi^2)^2 \phi^4
= \half \phi^4 - \phi^6 + \half \phi^8 \,.
\label{symmoct}
\ee
This has a quartic minimum at $\phi = 0$ with $\mu = 1$, and quadratic
minima at $\phi = \pm 1$. The kink of interest interpolates between
$\phi_- = 0$ and $\phi_+ = 1$, and a shift of $\phi$ is not
necessary. The kink tail is long-range on the left and short-range on
the right. Here, $U(\phi) = (1 - \phi^2) \phi^2$ and
\be
W(\phi) = \frac{1}{3}\phi^3 - \frac{1}{5}\phi^5 - \frac{2}{15} \,,
\ee
where we have set the constant of integration to $-\frac{2}{15}$,
so that $W(1) = 0$. The kink mass is $M = -W(0) = \frac{2}{15}$.  

The integral relation (\ref{formal}) for this potential takes the form
\be
\int^{\phi}_{} \left( \frac{1}{2(1 - \phi)} + \frac{1}{2(1 + \phi)} +
  \frac{1}{\phi^2} \right) d\phi = x \,,
\ee
from which follows the implicit kink solution \cite{Loh}
\be
\half\log\frac{1+\phi}{1-\phi} - \frac{1}{\phi} = x - {\cal{X}} \,.
\label{kink1}
\ee
${\cal{X}}$ we call the location of the kink, and it is the same
constant as in (\ref{quartkinktail}), as the left-hand kink tail derived from
(\ref{kink1}) is $\phi(x) = -\frac{1}{x - {\cal{X}}}$. The kink centre
is where $V$ (and $U$) has its maximum value. This is where
$\phi = \frac{1}{\sqrt{2}}$, and it occurs at
$X_0 = {\cal{X}} + \log(1 + \sqrt{2}) - \sqrt{2} \simeq {\cal{X}} - 0.533$. 
As $\mu = 1$, the antikink-kink force is therefore
\be
F = Ma = \half \frac{J^4}{{\cal{X}}^4}  =
\half \frac{J^4}{(X_0 + 0.533)^4} = \frac{1.477}{(X_0 + 0.533)^4} \,,
\label{longshortforce}
\ee
and in terms of the separation $s = 2X_0$,
\be
F = \frac{8 J^4}{(s + 1.066)^4} = \frac{23.63}{(s + 1.066)^4}\,.
\label{longshortsep}
\ee
This is a refinement of the result in ref.\cite{NM}. 

Despite the logarithmic term in (\ref{kink1}), there is no logarithmic
correction to the kink tail. The first correction is cubic in
$\frac{1}{x - \cal{X}}$. (A small quadratic correction
$\frac{\varepsilon}{(x - {\cal{X}})^2}$ could be interpreted as a
shift of $\cal{X}$ by $-\varepsilon$.) This is a consequence of the
symmetry of $V$, which implies that $V(\phi)$ has no $\phi^5$ term.
The leading tail behaviour is therefore unambiguous, and there is a
definite $O(1)$ difference between $X_0$ and $\cal{X}$. 

As a second example, consider the symmetric rational potential with two
quartic minima \cite{Ma6}
\be
V(\phi) = \half \frac{(1-\phi^2)^4}{(1+\phi^2)^2} \,.
\label{Vrat}
\ee
Its reflection-antisymmetric kink interpolates between $-1$ and $1$.
After a shift of $\phi$ by $1$, the potential is
\be
V(\phi) = \half \frac{(2-\phi)^4 \phi^4}{(2 - 2\phi + \phi^2)^2} \,.
\label{Vratshift}
\ee
The integral relation (\ref{formal}) simplifies to
\be
\half \int^{\phi}_{} \left( \frac{1}{(2 - \phi)^2}
+ \frac{1}{\phi^2} \right) \, d\phi = x \,,
\ee
which can be integrated to give
\be
\frac{1}{2 - \phi} - \frac{1}{\phi} = 2x \,,
\ee
and hence the explicit kink solution
\be
\phi_{\rm K}(x) = \frac{-1 + \sqrt{4x^2 + 1}}{2x} + 1 \,,
\ee
where the positive square root ensures that the kink interpolates between
$\phi_- = 0$ and $\phi_+ = 2$. This kink solution is centred at the
origin, as the potential has its maximum at $\phi = 1$, and
$\phi_{\rm K}(0) = 1$. The kink centred at $X_0$ is
\be
\phi_{\rm K}(x - X_0) = \frac{-1 + \sqrt{4(x-X_0)^2 + 1}}{2(x-X_0)}
+ 1 \,.
\ee

The expansion of the potential (\ref{Vratshift}) about $\phi = 0$ is
\be
V(\phi) = 2\phi^4 - \phi^6 + O(\phi^7) \,,
\label{Vratexpand}
\ee
so $\mu = 2$. $V(\phi)$ again has no quintic term (not in this case
because of a symmetry) and therefore the kink tail has no logarithmic
correction. Explicitly, the kink tail for $x \ll X_0$ has the expansion 
\be
\phi_{\rm K}(x-X_0) \simeq - \frac{1}{2(x-X_0)} - \frac{1}{8(x-X_0)^2}
+ O\left( \frac{1}{(x-X_0)^4} \right) \,,
\ee
which in turn can be expressed as
\be
\phi_{\rm K}(x-X_0) \simeq -\frac{1}{2\left(x-X_0-\frac{1}{4}\right)}
+ O\left( \frac{1}{(x-X_0)^3} \right) \,.
\ee
Therefore ${\cal{X}} = X_0 + \frac{1}{4}$, again a definite
$O(1)$ difference, and the antikink-kink force is
\be
F = Ma = \frac{1}{8} \frac{J^4}{\left(X_0 + \frac{1}{4}\right)^4} \,.
\ee
In terms of the separation $s = 2X_0$ the force is
\be
F = \frac{2 J^4}{\left( s + \half \right)^4} \,.
\ee

Our final example is the simplest octic potential with
two quartic minima
\be
V(\phi) = \half (1 - \phi)^4 \phi^4 \,,
\label{Vsimp}
\ee
whose reflection-antisymmetric kink interpolates between $0$ and
$1$ and whose centre $X_0$ is where $\phi = \half$. The expansion
of $V$ about $\phi = 0$ is
\be
V(\phi) = \half \phi^4 - 2\phi^5 + \cdots
\label{Vsimpexpand}
\ee
so $\mu = 1$. In this example, the quintic term leads to a logarithmic
correction to the kink's tail, as one can verify from the implicit
kink solution
\be
2\log\frac{\phi}{1 - \phi} + \frac{2\phi - 1}{(1 - \phi)\phi}
= x - X_0 \,.
\ee
We can therefore only determine the leading term in the antikink-kink
force. This is
\be
F = \frac{8J^4}{s^4} \,.
\ee
An improvement, taking account of the logarithmic term in the
kink tail, has recently been proposed and tested \cite{CM}. It would
be interesting to match this to a corresponding modification of
${\widetilde V}(\phi)$ and the integral (\ref{quartintegral}), to
include the effect of the quintic term in the potential. 

\section{The Kink Acceleration Equation}

We have recalculated the antikink-kink force using a uniform approach
that works for kinks with short-range or long-range tails. Several
examples, some previously known and some novel, have been discussed.
However, the calculations are not entirely satisfactory, even though
we have been careful to avoid known pitfalls.
First, the very notion of the antikink-kink force is rather fuzzy,
as it seems to depend on the instantaneous field configuration
and a definition of the kink centre. Second, it relies on several
approximations, and appears hard to make more rigorous, although
some informal considerations show that the errors are at subleading order.

A better, mathematically well-posed problem is suggested by
the accelerating kink equation,
\be
\frac{d^2\phi}{dx^2} - a\frac{d\phi}{dx} - \frac{dV(\phi)}{d\phi} = 0 \,,
\label{accelphi}
\ee
whose solution is defined on the half-line $x>0$ and is required to
satisfy the boundary conditions $\frac{d\phi}{dx} = 0$ at $x=0$
and $\phi(x) \to \phi_+$ as $x \to \infty$,
where $\phi_+$ is as before. The value $\phi = \phi_a > \phi_-$
at $x=0$ is not initially specified, and will depend on $a$. The
mechanical analogy \cite{Raj} that regards (\ref{accelphi}) as
a Newtonian dynamics with friction makes it plausible that
$\frac{d\phi}{dx} > 0$ for all $x > 0$, so the solution
is strictly monotonic.

Equation (\ref{accelphi}) has been numerically solved \cite{dOr} for the 
symmetric octic potential (\ref{symmoct}), verifying the monotonicity
of the solution. The main result is the relation between
the force $F=Ma$ and the location of the kink centre $X_0$. This
is presented graphically, but also as an
algebraic fit whose leading terms for large $X_0$ are
\be
F = \frac{1}{X_0^4} \left( 1.477 - \frac{3.1}{X_0} \right) \,.
\ee
This agrees well with the formula (\ref{longshortforce}), including
the O(1) shift of $\cal{X}$ relative to $X_0$. It would be worthwhile to
see if this approach is successful for other potentials with a quartic
minimum, both with and without logarithmic corrections to the kink tail. 

Equation (\ref{accelphi}) can be solved directly, as in ref.\cite{dOr}, but
it can alternatively be reduced to a first-order equation by
introducing a function $W_a(\phi)$ and making the ansatz
\be
\frac{d\phi}{dx} = \frac{dW_a(\phi)}{d\phi} \,.
\label{Wadefn}
\ee
Then $\frac{d^2\phi}{dx^2} = \frac{d^2W_a}{d\phi^2}\frac{dW_a}{d\phi}$ so
eq.(\ref{accelphi}) becomes
\be
\frac{d}{d\phi} \left\{ \half\left( \frac{dW_a}{d\phi} \right)^2
- aW_a(\phi) - V(\phi) \right\} = 0 \,,
\ee
which integrates to
\be
\half\left( \frac{dW_a}{d\phi} \right)^2 - aW_a(\phi) - V(\phi) = 0 \,.
\label{Waeq}
\ee
The function $W_a(\phi)$ is only defined up to an additive constant in 
(\ref{Wadefn}), so the constant of integration in (\ref{Waeq})
can be set to zero. Because of the monotonicity of $\phi(x)$, and
hence of $\frac{dW_a}{d\phi}$ as a function of $\phi$, this
equation for $W_a(\phi)$ has a single-valued solution, with
$\frac{dW_a}{d\phi} = 0$ at the endpoints $\phi_a$ and
$\phi_+$. Because $V(\phi_+) = 0$, it follows that $W_a(\phi_+) = 0$.

In the limit $a \to 0$, $W_a(\phi)$ becomes the function $W(\phi)$
that appears in eq.(\ref{Bogo}) for the unaccelerated kink,
which is often known explicitly. For general $a>0$, there appears to be
no explicit solution, and it is necessary to solve eq.(\ref{Waeq})
numerically to find $W_a(\phi)$ in the range $\phi_a \le \phi \le
\phi_+$. A shooting method is required to fix
the lower boundary $\phi_a$. Then, with $W_a(\phi)$ determined, one
would need to solve eq.(\ref{Wadefn}) numerically to find the
accelerating kink profile. Finally one would need
to determine the centre $X_0$ of the kink, the location where $V(\phi)$
has its local maximum. (The relevant value of $\phi$ is easily
found from $V(\phi)$, and is independent of $a$.) Combining these
results would give the graphical relation between the acceleration $a$
and the kink separation $s = 2X_0$. 

\section{Conclusions}

We have revisited the calculations of the force between a well-separated
antikink and kink. It has been shown that a uniform method using the
ansatz for a coherently accelerating kink gives good results for a
broad range of field theory potentials $V(\phi)$. The method applies to
potentials with quadratic minima, where the antikink-kink interaction
is short-range and falls off exponentially, and to those with
quartic minima, where the interaction is long-range and the force
falls off with the inverse fourth power of the separation. The
coefficients of the leading terms have been unambiguously determined.
Our calculations rely on an approximate treatment of the equation for
an accelerating kink; however, for one potential with a
quartic minimum, a more precise solution has confirmed the validity
of the approximation. Further work is needed to obtain a
detailed picture of the force beyond leading order when the potential has
a quartic minimum with a quintic subleading term, leading to a kink
tail with a logarithmic subleading term. 

\vspace{4mm}

\section*{Acknowledgements}

I am grateful to Andrzej Wereszczynski and Katarzyna S\l{}awi\'{n}ska
for their help initiating this research. This work has been partially
supported by STFC consolidated grant ST/P000681/1.

\end{document}